\documentclass[showpacs,preprintnumbers,amsmath,amssymb,color]{revtex4}

\usepackage[dvips]{color}
\usepackage{graphicx}
\usepackage{dcolumn}
\usepackage{bm}

\graphicspath{%
    {converted_graphics/}
    {C:/Users/Nail/Desktop/}
    {C:/Users/Nail/Desktop/Mathematica/}
}
\begin{document}

\preprint{IUHET-539}

\title{Second-order Supersymmetric Operators and Excited States}

\author{Micheal S. Berger}
 \email{berger@indiana.edu}
\author{Nail S. Ussembayev}%
 \email{nussemba@indiana.edu}
\affiliation{%
Department of Physics, Indiana University, Bloomington, IN 47405, USA}%

\begin{abstract}
 Factorization of quantum mechanical Hamiltonians has been a useful technique for some time. This 
procedure has been given an elegant description by supersymmetric quantum mechanics, and the subject has become well-developed. We demonstrate that the existence of raising and lowering operators 
for the harmonic oscillator (and many other potentials) can be extended to their supersymmetric 
partners. The use of double supersymmetry (or a factorization chain) is used to obtain non-singular isospectral potentials, and the explicit expressions for the ladder operators, wave functions and probability densities are provided. This application avoids the technical complexities of the most general approaches, and requires relatively modest methods from supersymmetric quantum mechanics.    
\end{abstract}
\pacs{03.65.Ge, 03.65.Fd, 03.65.Ca}

\maketitle

In supersymmetric quantum mechanics (SUSY QM) the form of the Schr\"odinger equation of a particle is determined by its Hamiltonian ($\hbar=2m=1$): \[H=-\partial_x^2+V(x)\] where $V(x)$ is nonsingular on $-\infty<x<\infty$ and has at least one bound state. If its ground state wave function $\psi_0$ and energy $E_0$ are known, then it is always possible to factor the Hamiltonian $H_-^{(0)}= H-E_0$ into a product of two linear differential operators: \[H_-^{(0)}=-\partial_x^2+V_-^{(0)}(x)=(-\partial_x+W_0)(\partial_x+W_0)=A^\dag_0 A_0\] where $W_0(x)$ is the superpotential related to the potential $V_-^{(0)}(x)$ by the Riccati equation: \[V_-^{(0)}(x)=W_0^2(x)-W_0'(x)=\psi_0''(x)/\psi_0(x).\] The superpotential can be written in terms of the ground state wave function as $W_0(x)=-\psi_0'(x)/\psi_0(x)$ with primes denoting the derivative. By reversing the order of $A_0^\dag$ and $A_0$, one obtains a new Hamiltonian \[H_+^{(0)}=A_0A_0^\dag=-\partial_x^2+V_+^{(0)}(x)=-\partial_x^2+W_0^2(x)+W_0'(x)\] whose eigenvalues and eigenfunctions are related to those of $H_-^{(0)}$. For instance, the eigenfunctions of $H_+^{(0)}$ $ (H_-^{(0)})$ can be found by applying the operator $A_0$ $ (A_0^\dag)$ to the eigenfunctions of $H_-^{(0)}$ $ (H_+^{(0)})$.  Moreover, it turns out that these partner Hamiltonians have identical spectra (except for the ground state of $H_+^{(0)}$), so that having an exactly soluble problem for $H_-^{(0)}$ guarantees derivation of the exact solution of $H_+^{(0)}$ \cite{Cooper}.

The generalization of this formalism to the excited states $\psi_n(x)$ $(n=1, 2, \cdots)$ of $H_-^{(0)}$ is straightforward \cite{Robnik}. It is done by introducing the operators \[A_n^\dag=-\partial_x+W_n(x) \mbox{ and } A_n=\partial_x+W_n(x)\] with superpotentials $W_n(x)=-{\psi_n'(x)/\psi_n(x)}$. Now we have two sets of potentials $V_\pm^{(n)}(x)=W_n^2(x)\pm W'_n(x)$ corresponding to the wave function $\psi_n(x)$. From the Schr\"odinger equation it follows that \[V_-^{(n)}(x)=V_-^{(0)}(x)+E_0-E_n\] (where $E_n$ is the energy eigenvalue corresponding to $\psi_n(x)$), so the potentials $V_-^{(n)}(x)$ offer nothing new. On the other hand, the potentials \[V_+^{(n)}(x)=2W_n^2(x)-V_-^{(n)}(x)\] are less trivial and more interesting.  A few remarks about this generalization are in order here. Since \[H_+^{(n)}[A_n\psi_k]=A_nA_n^\dag A_n\psi_k =A_n[H_-^{(n)}\psi_k]=(E_k-E_n)[A_n\psi_k]\] one may conclude that $A_n\psi_k$ is an eigenfunction of the Hamiltonian $H_+^{(n)}$ with energy eigenvalue $E_k-E_n$ (note that the normalization condition requires $k>n$). However, from the orthogonality of eigenfunctions $\psi_n(x)$ we know that an excited state wave function changes sign at one or more points in space and consequently the superpotentials $W_n(x)$ are always singular for $n>0$. Therefore, the proper behavior of the wave function $\psi_k(x)$ does not necessarily mean that $A_n\psi_k(x)=(\partial_x+W_n(x))\psi_k(x)$ is an acceptable solution, and the states of $H_-^{(n)}$ are not, in general, in one-to-one correspondence with the states of $H_+^{(n)}$ unless $n=0$, which corresponds to the standard unbroken SUSY. That is, for excited states the degeneracy of energy levels typically breaks down and this breakdown explains $\bold{relatively}$ low interest in extensions of SUSY QM to the excited states of a potential. As an illustration consider the following exercise. One dimensional harmonic oscillator eigenfunctions are given by $\psi_n(y)=H_n(y)e^{-y^2/2}$ where $H_n(y)$ are Hermite polynomials and $y=\sqrt{\omega/2} x$. The superpotentials are easily derived by using the well known recursion relations for Hermite polynomials: \[W_n(y)=\sqrt{\omega\over2}\left(\frac{H_{n+1}(y)}{H_n(y)}-y\right). \] They can be used to generate the first two pairs of partner potentials according to $V_\pm^{(n)}(x)=W_n^2(x)\pm W'_n(x)$: \[V_-^{(0)}(x)={\omega^2x^2\over4}-{\omega\over2} \mbox{ and } V_+^{(0)}(x)={\omega^2x^2\over4}+{\omega\over2}; \hspace{0.5cm}
V_-^{(1)}(x)={\omega^2x^2\over4}-{3\omega\over2} \mbox{ and } V_+^{(1)}(x)={\omega^2x^2\over4}-{\omega\over2}+{2\over x^2}.\]

The $n=0$ case is the standard isospectral SUSY QM (the lowest state of $V_-^{(0)}(x)$ has no counterpart in the spectra of $V_+^{(0)}(x)$), whereas the $n=1$ case corresponds to the generalization of SUSY QM to the first excited state and one expects the breakdown of the degeneracy of energy levels. Indeed, the potential $V_+^{(1)}(x)$ defined on the interval $[0,\infty)$ is known as isotonic oscillator  \cite{Weissman}. Its eigenfunctions \[A_1\psi_n(y)=\sqrt{\omega\over2}\left[{d\over dy}+y-{1\over y}\right]e^{-y^2/2}H_n(y)\] are only nonsingular for odd $n$ and consequently the energy level spacing is twice larger than that of its partner $V_-^{(1)}(x)$, $\bold{i.e.}$ the energy levels are partially degenerate. In the next pair of potentials $V_-^{(2)}(x)$ and $V_+^{(2)}(x)$ degeneracy is completely absent \cite{Panigrahi}.  Note that we investigate the degeneracy between the potentials $V_{\pm}^{(n)}$ for some fixed $n$ and not between two unrelated pairs of potentials $V_{\pm}^{(n)}$ and $V_{\pm}^{(n')}$.

It is clear that in the traditional factorization method (i.e. the first-order SUSY QM) there is no way to use the excited states of the initial potential and at the same time avoid creating singularities in the partner potential. The second-order SUSY QM can be used to overcome this difficulty \cite{Rosas}. It can be formulated by the following intertwining relationship \[\tilde H A=AH\] where two different Hamiltonians $\tilde H=-\partial_x^2+\tilde V(x)$ and $H-\partial_x^2+V(x)$ are intertwined by the operator of the second order in derivatives: \[A=\partial_x^2+\eta(x)\partial_x+\gamma(x).\]  There exist two kind of operators $A$  -- reducible and irreducible. If it is not possible to write $A$ as a product of two first order operators with real superpotentials, then they are referred to as irreducible \cite{Andrianov}. In this paper we generate families of strictly isospectral potentials by using reducible operators which are equivalent to two successive first-order intertwining relations with real superpotentials. In other words, the relationship between partner potentials in our framework is the same as that obtained by the confluent algorithm \cite{Fernandez}, but the novelty of our approach is  based on exploiting the non-uniqueness of  factorization in the construction of partner Hamiltonians \cite{Mitra}. In this case there is an atypical Hamiltonian $\tilde H_+^{(n)}$ (singular for $n>0$) that admits non-unique factorization.  This Hamiltonian plays an intermediate role in the construction and we do not discuss the degeneracy of its energy levels  -- we merely use it to construct two isospectral potentials. Thus, we succeed in avoiding some common obstacles (such as singular potentials, failure of the degeneracy theorem that arises when dealing with excited states etc.) by using non-uniqueness of factorization at the intermediate stage. 
 Continuing our exercise with the harmonic oscillator we show below how to construct two one-parameter class of potentials which are not the harmonic oscillator, yet have the same energy spectrum. We explicitly derive the associated wave functions and plot the probability densities.

The starting point is to define the operators 

\[B_{n}=\partial_x+f(x)+W_n(x) \mbox{ and } 
B^\dag_{n}=-\partial_x+f(x)+W_n(x)\] where $f(x)$ is temporarily undetermined function. 
As usual we demand that $\tilde H_+^{(n)}=B_{ n}B^\dag_{ n} $ which results in \[\tilde H_+^{(n)}=-\partial_x^2+V_+^{(n)}(x) +f'(x)+2W_n(x)f(x)+f^2(x).\]  In order to arrive at the non-unique factorization $\tilde H_+^{(n)}=H_+^{(n)}$ one should set \[f'(x)+2W_n(x)f(x)+f^2(x)=0\] and solve for $f(x)$. This differential equation is known as the Bernoulli equation (a specific example of the Riccati 
equation) and it possesses the following general solution which depends on an arbitrary integration constant $C$ that can be considered as a free parameter: \[f_n(x)=\frac{\psi_n^2(x)}{C +\int_{x_0}^x{\psi_n^2(s)ds}}\] where $C$ and $x_0$ are constants. A little calculation shows that the partner Hamiltonian $\tilde H_-^{(n)}=B^\dag_{ n}B_{ n}$ is given by \[\tilde H_-^{(n)}=-\partial_x^2+V_-^{(n)} -2f_n'(x)=-\partial_x^2+\tilde V_-^{(n)}(x).\] 

Obviously, the potential $V_+^{(n)}(x)$ is free of singularities for $n=0$ and it is legitimate to discuss the degeneracy of energy levels of the potentials $V_-^{(0)}(x)$, $V_+^{(0)}(x)$ and $\tilde V_-^{(0)}(x)$ (the latter is assumed to be nonsingular for certain values of the parameter $C$). However, as it was pointed out earlier for the excited states the potential $V_+^{(n)}(x)$ has singularities corresponding to the zeros of the wave function $\psi_n(x)$. So while the potential $V_-^{(n)}(x)$ is defined on the full real axis, the range of $V_+^{(n)}(x)$ is composed of $n+1$ disjoint regions and their energy spectra are non-degenerate or only partially degenerate as explained above. An interesting question arises naturally: is the degeneracy theorem still valid for the potentials $V_-^{(n)}(x)$ and $\tilde V_-^{(n)}(x)$ under the assumption that $\tilde V_-^{(n)}(x)$ is free of singularities for certain $C$? We have found that the intermediate (or atypical) Hamiltonian $H_+^{(n)}$ admits two different factorizations due to its non-uniqueness, namely, $A_nA_n^\dag=B_{n}B_{n}^\dag$. Hence, we can exploit the operators $A_n$ and $B_{n}$ together with their adjoint to answer the above question. 

The Schr\"odinger equation $H_-^{(n)}\psi_k=(E_k-E_n)\psi_k$ implies \[\tilde H_-^{(n)}[B_{n}^\dag A_n\psi_k]=B_{n}^\dag B_{n}B_{n}^\dag A_n\psi_k = B_{n}^\dag A_nA_n^\dag A_n\psi_k=(E_k-E_n)[B_{n}^\dag A_n\psi_k]\]where $k\ne n$,
i.e. if $\psi_k(x)$ is an eigenfunction of the Hamiltonian $H_-^{(n)}$ with energy eigenvalue $E_k-E_n$, then $B_{n}^\dag A_n\psi_k$ is an eigenfunction of $\tilde H_-^{(n)}$ with the same energy. Similarly, from the Schr\"odinger equation $\tilde H_-^{(n)}\tilde\psi_k^{(n)}=\tilde E_k^{(n)}\tilde\psi_k^{(n)}$ (where in $\tilde E_k^{(n)}$, $k$ denotes the energy level and $(n)$ refers to the $n^{th}$ eigenfunction of the Hamiltonian $H_-^{(0)}$) it follows that 

\[H_-^{(n)}[A_n^\dag B_{n}\tilde\psi_k^{(n)}]=\tilde E_k^{(n)}[A_n^\dag B_{n}\tilde\psi_k^{(n)}]\]

It is clear that the normalized eigenfunctions of the Hamiltonians $H_-^{(n)}$ and $\tilde H_-^{(n)}$ are related by \[\tilde\psi_k^{(n)}(x)=(E_k-E_n)^{-1}[ B_{n}^\dag A_n\psi_k(x)] \mbox{ and } \psi_k(x)=(E_k-E_n)^{-1}[ A_n^\dag B_{n} \tilde\psi_k^{(n)}(x)]\] with $k\ne n$. The operators $A_n$ or $B_{n}$ destroy a node in the eigenfunctions, but they are followed respectively by the operators $B_{n}^\dag$ or $A_n^\dag$ that create an extra node. So the overall number of the nodes does not change. In addition, the normalization does not require positive semi-definiteness of the energy eigenvalues, as in the standard case. The operator $B_{n}^\dag A_n$ can be factored and compared with the above mentioned operator $A=\partial_x^2+\eta(x)\partial_x+\gamma(x)$. One realizes that $-\eta(x)=f_n(x)$ and $-\gamma(x)=V_-^{(n)}(x)+f_n(x)W_n(x)$. 

   For any $n$ there is always one missing state $k=n$ (since $A_n\psi_n(x)=0$) which can be obtained by solving the first order differential equation $B_{n}\tilde\psi_n^{(n)}=0$:
\[{d\tilde\psi_n^{(n)}(x)\over dx}=-\left(W_n(x)+\frac{\psi_n^2(x)}{C +\int_{x_0}^x{\psi_n^2(s)ds}}\right) \tilde\psi_n^{(n)}(x)={d\over dx}\left(\ln\frac{\psi_n}{C +\int_{x_0}^x{\psi_n^2(s)ds}}\right)\tilde\psi_n^{(n)}(x)\]
In other words, \[\tilde \psi_n^{(n)}(x)=N(C)\times\frac{\psi_n}{C +\int_{x_0}^x{\psi_n^2(s)ds}}\]
with the corresponding energy $\tilde E_n^{(n)}=0$. All other energy eigenvalues satisfy $\tilde E_k^{(n)}=E_k-E_n$. Negative energy states appear when $n>0$. The normalization constant $N(C)$ depends on the parameter $C$ and other parameters of the initial potential $V_-^{(n)}(x)$. It is a constraint that allows one to determine the values of $C$ for which the potentials $\tilde V_-^{(n)}(x)$  are nonsingular and eigenfunctions $\tilde\psi_k^{(n)}(x)$ are well-defined.

Let us apply the developed theory to the harmonic oscillator potential (with $\omega=2$) whose wave functions $\psi_k(x)=H_k(x)e^{-x^2/2}$ and energy eigenvalues $E_k=2k$ are well known. 
 Considering first the $n=0$ case, one obtains the following  partner potentials \[V_-^{(0)}(x)=x^2-1 \mbox{ and }  \tilde V_-^{(0)}(x)=x^2-1-2{d\over dx}\left(\frac{e^{-x^2}}{C+\int_0^x{e^{-s^2}ds}}\right).\] The latter is free of singularities when $|C|>\sqrt \pi/2$ (as follows from the normalization constant $N(C)$) and asymptotically behaves like $x^2-1$ (see Fig. 1).

\begin{figure}[ht] 
\begin{minipage}[b]{0.45\linewidth}
\centering
  \includegraphics[bb=0 0 240 237,width=3.2in,height=3.2in]{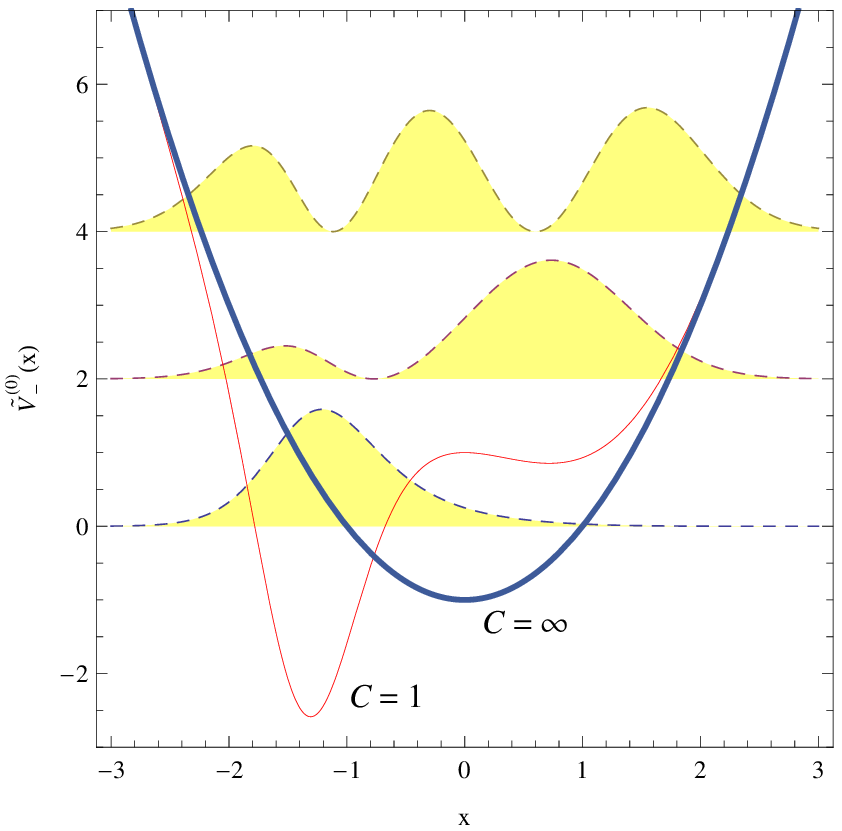}
  \caption{Plot of potential $\tilde V_-^{(0)}(x)$ with $C=1$ and unnormalized probability densities (dashed line at the corresponding level position) for its three lowest energy eigenvalues. The limit $C\to\infty$ corresponds to the potential $V_-^{(0)}(x)=x^2-1$ (thick blue line).}
  \label{fig:CorrectedPaperFig1}
\end{minipage}
~\hfill~
\begin{minipage}[b]{0.45\linewidth}
\centering
    \includegraphics[bb=0 0 240 236,width=3.2in,height=3.2in]{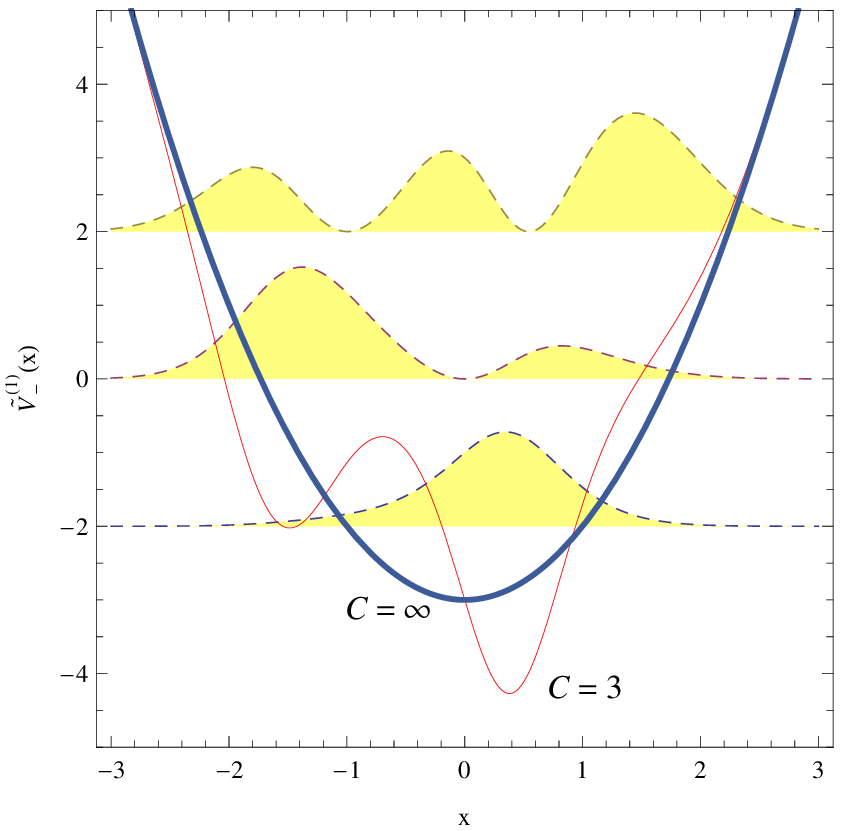}
  \caption{Plot of potential $\tilde V_-^{(1)}(x)$ with $C=3$ and unnormalized probability densities (dashed line at the corresponding level position) for its three lowest energy eigenvalues. The limit $C\to\infty$ corresponds to the potential $V_-^{(1)}(x)=x^2-3$ (thick blue line).}
  \label{fig:CorrectedPaperFig2}
\end{minipage}
\end{figure}
Its eigenvalues are $\tilde E_k^{(0)}=2k$, that is identical to those of the harmonic oscillator and the normalized ground state wave function reads \[\tilde\psi_0^{(0)}(x)=\sqrt{4C^2-\pi\over{4\sqrt\pi}}\frac{e^{-x^2/2}}{C +\int_{x_0}^x{e^{-s^2}ds}}.\] All other wave functions are derived through $B_{0}^\dag A_0\psi_k$ or by applying  the fifth order raising operator  $(B_{0}^\dag A_0) A_0^\dag (A_0^\dag B_{0})$ repeatedly to the normalized first excited state wave function of the potential $\tilde V_-^{(0)}(x)$: \[\tilde\psi_1^{(0)}(x)=\left(4\over\pi\right)^{1/4}\frac{e^{-3x^2/2}(1+2 C x e^{x^2}+\sqrt\pi x\operatorname{erf}(x)e^{x^2})}{2 C+\sqrt\pi\operatorname{erf}(x)}.\]

The same result was obtained by Mielnik \cite{Mielnik} and using the Gelfand-Levitan formalism by Abraham and Moses \cite{Abraham}. Mielnik pointed out the third order raising operator for this problem $b^\dag a^\dag b$ where $b$ and $b^\dag$ convert the eigenfunctions of the partner Hamiltonians  into one another and $a^\dag$ is the usual creation operator. We immediately recognize that $b=A_0^\dag B_{0}$ and $a^\dag=A_0^\dag$ -- this is just the consequence of non-uniqueness of factorization (and in the first instance of double supersymmetry). Being able to reproduce this result is noteworthy, but another important fact is that one can obtain infinitely many such potentials (since the harmonic oscillator has infinitely many wave functions) and Mielnik's result is just one of them, namely $n=0$. We show one more example and then turn to the most general case. This time letting $n=1$ we find a pair of partner potentials: \[V_-^{(1)}(x)=x^2-3 \mbox{ and } \tilde V_-^{(1)}(x)=x^2-3-8{d\over dx}\left(\frac{x^2e^{-x^2}}{C-2xe^{-x^2}+\sqrt\pi \operatorname{erf}(x)}\right).\] The potential $\tilde V_-^{(1)}(x)$ is nonsingular for $|C|>\sqrt\pi$ (as follows from normalizing $\psi_1^{(1)}(x)$) and behaves like $x^2-3$ for large $x$ (see Fig. 2). Its eigenvalues are $\tilde E_k^{(1)}=2(k-1)$, i.e. the same as for $V_-^{(1)}(x)$ and there is one negative energy state as expected since we used the first excited state of the original potential. The normalized first excited state of the potential $\tilde V_-^{(1)}(x)$ corresponds to the energy $\tilde E_1^{(1)}=0$ and satisfies $B_1\tilde\psi_1^{(1)}(x)=0$: \[\tilde\psi_1^{(1)}(x)=\sqrt{C^2-\pi\over{2\sqrt\pi}}\frac{2x e^{-x^2/2}}{C-2xe^{-x^2}+\sqrt\pi\operatorname{erf}(x)}.\]  The normalized lowest state wave function is given by \[\tilde\psi_0^{(1)}(x)=\left(1\over\pi\right)^{1/4}\frac{e^{x^2/2}(C+\sqrt\pi\operatorname{erf}(x))}{Ce^{x^2}-2x+\sqrt\pi\operatorname{erf}(x)e^{x^2}}.\] Other states $k=2,3,\cdots$ are obtained by the usual construction or by applying the raising operator $(B_{1}^\dag A_1) A_0^\dag (A_1^\dag B_{1})$ to the wave function $\tilde\psi_2^{(1)}(x)$.

At least in the case of the harmonic oscillator one can provide explicit expressions for the pairs of isospectral potentials generated by an arbitrary excited state $\psi_n(x)$. To this end note that we need to evaluate the integral (with $x_0=0$) that appears in the denominator of the function $f_n(x)$: \begin{eqnarray*}\int_0^x \psi_n^2(s)ds=\int_0^x e^{-s^2}H_n^2(s)ds&=&\int_0^x ds e^{-s^2}\sum_{m=0}^n\frac{2^m}{m!}\left[\frac{n!}{(n-m)!}\right]^2 H_{2(n-m)}(s)\\&=&x\ (n!)^2\sum_{m=0}^n\frac{(-1)^m\ 2^{n-m}\ (2m)!}{(n-m)!\ (m!)^3}\ {}_1F_1\left(m+{1\over 2};{3\over 2};-x^2\right)\end{eqnarray*}  
where ${}_1F_1(\alpha;\beta;z)$ is the confluent hypergeometric function (cf. eq. (46) in ref. \cite{Fernandez}). 

Up until now we were assuming that for certain values of the parameter $C$ the function $f_n(x)$ is free of singularities. Of course, $f_n(x)$ is non-singular if its denominator is nodeless and latter increases monotonically (the derivative of the denominator is positive for all $x$). So the requirement is satisfied if the limits \[\lim_{x\to\pm\infty}\left\{C+\int_0^x \psi_n^2(s)ds\right\}\] both have the same sign, i.e. when \[|C|>\int_0^\infty \psi_n^2(s)ds\] and this is true for normalizable wave functions of any potential. In our case we find $|C|>2^{n-1}n!  \sqrt\pi$ which is in agreement with the above considered particular cases $n=0$ and $n=1$. Thus, one obtains  non-singular potentials $\tilde V_-^{(n)}(x)$ strictly isospectral to the harmonic oscillator $V_-^{(n)}=x^2-(2n+1)$. The same result (up to the choice of the energy scale) was explored for the first time in ref. \cite{Fernandez}. We would like to stress that for a given $n$ we supersymmetrize the original potential by subtracting the appropriate energy.

Continuing our theme of potentials with the oscillator spectrum let us finally apply the presented method to the exactly solvable non-polynomial one-dimensional quantum potential discovered recently by Cari\~nena, Perelomov, Ra\~nada and Santander (CPRS) \cite{CPRS}. In their original manuscript the authors solved the Schr\"rodinger equation for the potential \[V_{CRPS}(x)=\frac{x^2}{2}+4\frac{2x^2-1}{(2x^2+1)^2}\] assuming that the solution can be expressed as a power series in $x$ and deriving the recursion formula which determines the coefficients of the expansion. Later a much more simple solution was obtained by Fellows and Smith \cite{Fellows} who showed that the potential $V_-^{(0)}(x)=2V_{CPRS}+3$ is a supersymmetric partner of the harmonic oscillator $x^2+5$. We further analyze the CPRS potential with the aim to find new potentials with the spectrum of a simple oscillator. The energy eigenvalues and unnormalized wave functions of the potential $V_-^{(0)}(x) $ are given by $E_k=2k$ and \[\psi_k(x)=\frac{H_k(x)+4k H_{k-2}(x)+4k(k-3)H_{k-4}(x)}{2x^2+1}\ e^{-x^2/2}=\frac{P_k(x)}{2x^2+1}\ e^{-x^2/2}\] with $k=0,3,4,5,\dots$   Knowing one of the excited state wave functions, say, \[\psi_3(x)=\frac{4x(3+2x^2)}{2x^2+1}e^{-x^2/2}\] one can generate a pair of potentials with identical spectra (see Fig. 3):

\begin{figure}[h] 
  \centering
  \includegraphics[bb=0 0 240 240,width=3.2in,height=3.2in,keepaspectratio]{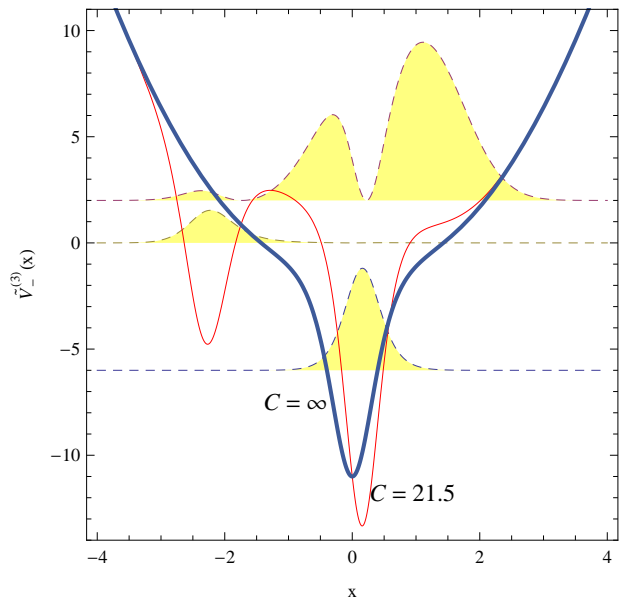}
  \caption{Plot of the potential $\tilde V_-^{(3)}(x)$ with $C=21.5$ (close to $12\sqrt\pi$) and unnormalized probability densities (dashed line at the corresponding level position) for its three lowest energy eigenvalues. The limit $C\to\infty$ corresponds to the CPRS potential $V_-^{(3)}(x)$ (thick blue line).}
  \label{Fig.3:}
\end{figure}

 \[V_-^{(3)}(x)=x^2-3+8\frac{2x^2-1}{(2x^2+1)^2} \mbox{ and } \tilde V_-^{(3)}(x)=V_-^{(3)}(x)+2{d\over dx}\left(\frac{16 x^2 (2x^2+3)^2}{8x(4x^4+8x^2+3)-e^{x^2}(2x^2+1)^2(C+12\sqrt\pi \operatorname{erf}(x))}\right)\]  The latter is free of singularities for $|C|>12\sqrt\pi$. It might seem very unexpected that this highly non-trivial potential possesses such a simple spectrum as $\tilde E_k^{(3)}=2k-6$ for $k=0,3,4,5,\dots$ Its first excited state wave function reads \[\tilde\psi_3^{(3)}(x)=\sqrt{\frac{C^2-144\pi}{24\sqrt\pi}}\frac{4x(2x^2+3)e^{x^2/2}}{e^{x^2}(2x^2+1)(C+12\sqrt\pi\operatorname{erf}(x))-8x(2x^2+3)}.\] The eigenfunctions with $k\ne n$ are obtained by applying the operator $B_n^\dag A_n$ to the wave functions $\psi_k(x)$ of the original potential $V_-^{(0)}(x)$.

It turns out that for this problem too one can write the partner potentials generated by an arbitrary excited state $\psi_n(x)$. We need the following two identities which are valid for any $n\ge3$ (proofs can be found in \cite{CPRS}): \[P_n'(x)=4n(2x^2+1)H_{n-3}(x) \mbox{ and } \frac{P_n(x)e^{-x^2}}{(2x^2+1)^2}=-2{d\over dx}\left(\frac{H_{n-3}(x)e^{-x^2}}{2x^2+1}\right).\] Using these identities and integrating by parts we find \begin{eqnarray*}\int_0^x{\psi_n^2(s)ds}&=&-\frac{2P_n(x)H_{n-3}(x)e^{-x^2}}{2x^2+1}+8n\int_0^x{e^{-s^2}H_{n-3}^2(s)ds}\\&=&-\frac{2P_n(x)H_{n-3}(x)e^{-x^2}}{2x^2+1}+x\ 8n((n-3)!)^2\sum_{m=0}^{n-3}\frac{(-1)^m\ 2^{n-m-3}\ (2m)!}{(n-m-3)!\ (m!)^3}\ {}_1F_1\left(m+{1\over 2};{3\over 2};-x^2\right).\end{eqnarray*} Hence for any given $n$, the potentials \[V_-^{(n)}(x)=\frac{\psi_n''(x)}{\psi_n(x)}=x^2+8\frac{2x^2-1}{(2x^2+1)^2}-2n+3 \mbox{ and } \tilde V_-^{(n)}(x)=V_-^{(n)}(x)-2f_n'(x)\] are strictly isospectral with energy eigenvalues $2(k-n)$ where $k,n=0,3,4,5,\dots$. The potentials $\tilde V_-^{(n)}(x)$ are non-singular for $|C|>(n-3)!2^{n-1}n\sqrt\pi$ with $n\ge3$. We calculated $V_-^{(n)}(x)$ using the following identity which does not appear in ref. \cite{CPRS} where some properties of the polynomials $P_n(x)$ have been studied: \[\frac{2x\ (2x^2+5)}{2x^2+1}\frac{P_n'(x)}{P_n(x)}-\frac{P_n''(x)}{P_n(x)}=2n.\]  The canonical form for this identity is the Schr\"odinger equation for the CPRS potential.

The above discussion illustrates a general method for constructing isospectral potentials in the second-order SUSY QM. Given the wave functions $\psi_k$ and energy eigenvalues $E_k-E_0$  of the Hamiltonian $H_-^{(0)}$ we automatically get the wave functions $\psi_k$ and energy eigenvalues $E_k-E_n$ of the Hamiltonians $H_-^{(n)}$. This allows one to construct the Hamiltonian $\tilde H_-^{(n)}$ whose energy spectra is identical to that of $H_-^{(n)}$ and eigenfunctions are given by $\tilde\psi_k^{(n)}(x)\propto B_{n}^\dag A_n\psi_k(x)$ for $k\ne n$. Intertwining operators of higher order in derivatives have been studied \cite{Ioffe}. For the third-order reducible operator there will be two atypical Hamiltonians arising from the non-uniqueness of factorization. If the possibility of using operators is available (as when one uses excited states), then the operators inherit the property of being ladder operators at each stage.  

We presented simple examples to show the technique of constructing isospectral potentials,  and we have tested our results on the consistency with other methods. It is well-known \cite{Dong} that ladder operators can be constructed for other 
potentials such as Morse, P\"oschl-Teller, and the infinite square well. The raising and lowering operators for the double-supersymmetric partner potentials can be obtained along the same lines as 
illustrated in this paper. This avoids the unnecessarily technical challenges in the general problem,
and these potentials will be the subject of future investigations.

\section*{Acknowledgments}
N.U. was assisted by the Hutton Honors College Research Grant. M.B. was supported in part by the U.S.
Department of Energy under Grant No.~DE-FG02-91ER40661. 

\end{document}